\documentclass[amssymb,twocolumn,prb,superscriptaddress,floats,showkeys,showpacs,a4paper]{revtex4-1}

\usepackage[T1]{fontenc}
\usepackage{bm}
\usepackage{graphicx}
\usepackage{natbib}
\usepackage{epstopdf}
\usepackage[euler]{textgreek}
\usepackage{ulem}
\usepackage{float}
\usepackage{color}

\begin{document}

\title{The excited spin-triplet state of a charged exciton in quantum dots}

\author{M. R. Molas}
\email{maciej.molas@gmail.com}
\affiliation{Faculty of Physics, University of Warsaw, ul. Pasteura 5, 02-093 Warszawa, Poland}
\affiliation{Laboratoire National des Champs Magn\'etiques Intenses, CNRS-UGA-UPS-INSA-EMFL, 25, avenue des Martyrs, 38042 Grenoble, France}
\author{A. A. L. Nicolet}
\affiliation{Laboratoire National des Champs Magn\'etiques Intenses, CNRS-UGA-UPS-INSA-EMFL, 25, avenue des Martyrs, 38042 Grenoble, France}
\author{B. Pi\k{e}tka}
\affiliation{Faculty of Physics, University of Warsaw, ul. Pasteura 5, 02-093 Warszawa, Poland}
\author{A. Babi\'nski}
\affiliation{Faculty of Physics, University of Warsaw, ul. Pasteura 5, 02-093 Warszawa, Poland}
\author{M. Potemski}
\affiliation{Laboratoire National des Champs Magn\'etiques Intenses, CNRS-UGA-UPS-INSA-EMFL, 25, avenue des Martyrs, 38042 Grenoble, France}

\date{\today}

\begin{abstract}

We report on single-object spectroscopic studies of a characteristic set of three resonances which appear in GaAlAs/AlAs quantum dot structures. 
The experiments included the photoluminescence (PL) and photoluminescence excitation (PLE) spectroscopy, photon-correlation measurements, analysis of the linear polarization of the emission spectra as well as the PL and PLE spectroscopy performed as a function of magnetic field. 
The investigated resonances are assigned to three different states of a positively charged exciton: a singlet ground state, a conventional triplet state (involving an electron from the ground state electronic $s$-shell), and additionally, a highly excited triplet state (involving the electron from the excited electronic $p$-shell). 
Advantageously (due to partially suppressed relaxation efficiency), the emission due to the highly excited triplet can seen in the spectra which allows us to scrupulously determine its properties and compare them to those of the conventional triplet state. 
Whereas both triplet states show similar patterns of the fine structure and Zeeman splitting, the splitting amplitudes are found to be significantly different for these two states. 
The presented results emphasize the role of the symmetry of the electronic state on the properties of the triplet states of the two holes + electron excitonic complex.

\end{abstract}

\maketitle

%XXXXXXXXXXXXXXXXXXXXXXXX        INTRO
\section{Introduction \label{sec:Intro}}

Semiconductor quantum dots (QDs) provide quantum confinement to carriers of charge, \cite{bayernature} which results in a multitude of possible strongly interacting few-particle complexes.\cite{finley} Their energy structure reflects both the Coulomb interactions between carriers and the potential landscape of the confinement. Various complexes composed of electrons and holes were thoroughly investigated in optical experiments.\cite{dekel,hawrylak,hartmann,santori,benny2012} 
Those primarily include single electron-hole pairs (excitons) and single charged excitons (an electron-hole pair + additional electron or hole) but also complexes involving larger number of carriers. So far, however, the interest was largely focused on emission lines associated with recombination processes between ground levels of the conduction and valence bands, the $s$-shells. Much less attention has been paid to the higher spectral band, related to the $p$-shell levels of QDs, which appears several meVs above the $s$-shell band. \cite{arashidatri,molascharged}

In this work we present the optical spectroscopic study of a positively charged exciton (two holes + one electron) confined in a single GaAlAs dot. Three distinct spectral lines, recalled as X$^+$, X$^{+*}$, and X$^{+*}_p$, associated with this complex are identified, see Fig. \ref{fig:photopol}(a). X$^+$ is due to the usual ground singlet state, ($e_s h_s h_s$), composed of an $s$-shell electron and two $s$-shell holes with opposite spins. X$^{+*}$ and X$^{+*}_p$ lines are due to excited states of this three particle complex. The former, X$^{+*}$ line is attributed to a triplet state, ($e_s h_s h_p$), built of an $s$-shell electron and two holes, one from the $s$- and the second one from the $p$-shell. The latter, X$^{+*}_p$ line is argued to be due to an excited triplet state ($e_p h_s h_p$) composed of two, $s$- and $p$-shell holes and an electron from the $p$-shell. While the observation of X$^+$ as well as X$^{+*}$ lines is well documented in literature\cite{poem,benny2012,arashidatri,molascharged}, the properties of the X$^{+*}_p$ resonance are, to the best of our knowledge, reported for the first time.

All three states of the charged exciton give rise to the characteristic emission lines, which can be observed at low temperature even with the lowest excitation power density. This implies a partially suppressed relaxation of the excited X$^{+*}$ and X$^{+*}_p$ states towards the X$^+$ ground state. The fact is particularly intriguing with respect to the observation of the X$^{+*}_p$ emission which involves the recombination of $p$-shell electrons and is therefore well separated in energy from the X$^+$ and X$^{+*}$ lines. The identification of X$^+$, X$^{+*}$, and X$^{+*}_p$ emission lines as due to different configurations of the same complex is supported by photon cross-correlation diagrams which display the antibunching dip for each pair of the investigated lines. 
As expected, the fine structure splitting, which is well pronounced for the X$^{+*}$ is here reported to be characteristic of the X$^{+*}_p$ line as well. 
This is in accordance with a triplet character of both X$^{+*}$ and X$^{+*}_p$ states. Relaxation processes from the X$^{+*}_p$ towards lower energy states are not fully suppressed and the X$^{+*}_p$ resonance is seen in both the photoluminescence (PL) and photoluminescence excitation (PLE) spectra detected on the X$^{+*}$ and X$^+$ lines. Notably, the coinciding emission and PLE-lines due to X$^{+*}_p$ resonance can be well followed in both the PL and PLE spectra measured as a function of magnetic field. Alike the X$^{+*}$ state, the X$^{+*}_p$ resonance splits in magnetic field into four components which form two doublets, well separated in energy. Whereas both X$^{+*}$ and X$^{+*}_p$ resonances display a number of similar properties the amplitudes of the observed effects are quite distinct: the fine structure splitting is reduced but the energy separation between magnetically split doublets is enhanced for the X$^{+*}_p$ state as compared to the X$^{+*}$ state. Our experimental observations call for thorough theoretical studies to compare the properties of two triplet states of a positively charged exciton, one involving an $s$- and the second one the $p$-shell symmetry electron. Possible reasons for partially suppressed relaxation between different states of positively charged excitons are discussed.

%XXXXXXXXXXXXXXXXXXXXXXXX      PROCEDURE
\section{Sample and Experimental Setups \label{sec:procedure}}

The active part of the investigated structure was intentionally designed as a GaAs/AlAs type-II bilayer.\cite{truby,wysmolekE} Previous research showed that the bilayer is not perfect in the lateral directions: the Ga-rich inclusions, which can be seen as islands of $\textrm{Ga}_{1-x}\textrm{Al}_{x}\textrm{As}$  ($x$ < 0.33) replacing the original bilayer were formed. These islands show all attributes of typical QDs. They are characterized by relatively strong confined and remarkably low surface density (at the level of $10^5-10^6$ $\textrm{cm}^{-2}$). Their emission spectra are dispersed in a wide energy range, 1.56-1.68~eV, and their optical response is overall very similar despite the dots are characterised by different shapes/size and different chemical composition (Al content).\cite{wysmolekappa,molasnatural,martin,pietkaprb,molasphd}

Single dot spectroscopy (PL and PLE experiments) were carried out at liquid helium temperature using two different experimental setups for experiments in the absence and in the presence of magnetic fields. In either case, to excie the PL spectra, a tunable Ti:Sapphire laser was set at $\lambda$= 725~nm to assure the quasi-resonant excitation conditions, $i.e.$, to inject the electron-hole ($eh$) pairs directly into QDs.\cite{kazimierczuknon} The PLE signal was detected at the emission lines associated with the recombination from an $s$-shell when sweeping the excitation energy of the laser.

In experiments at zero magnetic field, the sample was placed on a cold finger of a continuous flow cryostat equipped with a short-distance optical access. A microscope objective was used to focus the excitation beam and to collect the emitted light within a 1~\textmu m$^2$ spot. The emission was detected by 0.5 m-long monochromators equipped with the charge-couple-device cameras (CCDs) and avalanche photodiodes. The linearly polarized emission spectra were measured using a motorised, rotating half-wave plate combined with a fixed linear polariser, placed in front of the spectrometers.\cite{molasjap} Photon correlation experiments were implemented in the Hanbury-Brown and Twiss configuration with additional spectral filtering.\cite{twist}

The second setup was dedicated to experiments performed in magnetic fields up to 14~T in the Faraday configuration. The sample was located on top of an x-y-z piezo-stage immersed in a helium bath. The laser light was coupled to one branch of the Y-shaped fiber, focused by the microscope objective (spot size around 1~\textmu m$^2$), and the signal was detected from the second branch of the fiber, by a 0.5~m-long monochromator equipped with a CCD camera.\cite{adamphysE}

%XXXXXXXXXXXXXXXXXXXXXXXX        RESULTS
\section{Experimental results and discussion \label{sec:results}}

\begin{figure}[b!]
\centering
\includegraphics[width=80mm]{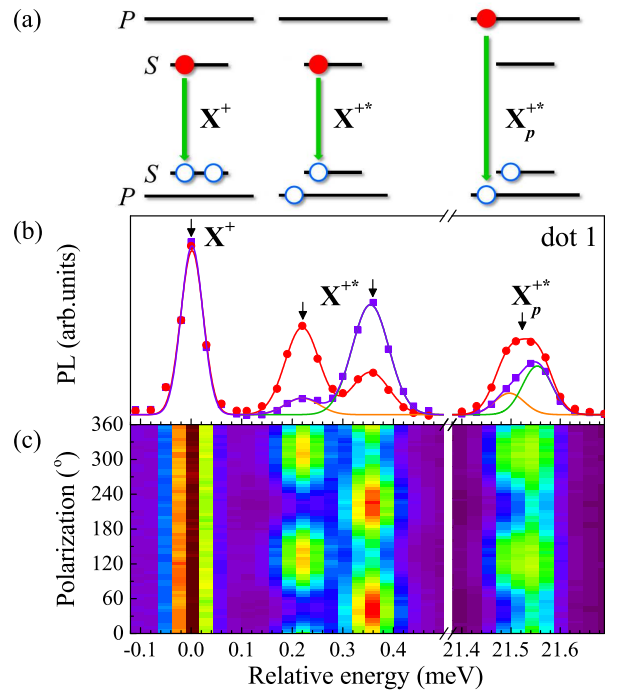}
\caption{(color online) (a) The schematic illustrations of transitions ascribed to the X$^+$, X$^{+*}$, and X$^{+*}_p$ emission lines. Red full and blue open circles represent electrons and holes on the $s$- and $p$-shell levels in the conduction and valence bands, respectively. Green vertical arrows indicate the associated recombination processes. (b) The PL spectra of a single GaAlAs QD measured at low excitation power and recorded for two perpendicular linear polarizations oriented along the crystallographic directions $[110]$ (violet points) and $[1\bar{1}0]$ (red points), respectively. The corresponding cumulative fits are drawn with the solid violet and red curves. The solid green and orange curves show a fit of the Gaussian function to the two split components of the X$^{+*}$ and X$^{+*}_p$ lines for the PL spectrum determined at a specific polarization. (c) Color plot presenting the PL spectra of the same QD measured using different linear polarizations of detection. The 45$^{\circ}$ and 135$^{\circ}$ correspond to the crystallographic axes $[110]$ and $[1\bar{1}0]$. The horizontal scale is shifted with reference to the energy of the X$^+$ line ($\mathrm{E}_{\mathrm{X}^{\mathrm{+}}}$=1.60615~eV).
\label{fig:photopol}}
\end{figure}

Emission spectra of collected from the investigated QDs show a number of sharp lines even when assuring a possible low level of the excitation power, which is typical for a majority of QD systems. More than 20 single QDs from the investigated sample were tested. The emission from each dot appears in slightly different spectral range and the relative intensity of varied emission lines changes from dot to dot. However the characteristic pattern of the emission lines can be always well recognized. With extensive analysis of linearly polarized emission from the investigated dots and photon correlation measurements, it was possible to distinguish three groups of lines: those related to (i) neutral, (ii) negatively-charged, and (iii) positively-charged excitons.\cite{molascharged,molasphd,molasepl} Here we focus our attention on a selected set of three lines which are assigned,at this point hypothetically, to three different states of a positively charged exciton: its singlet ground state (X$^+$) and two excited triplet states (X$^{+*}$, and X$^{+*}_p$). This assumption is verified in the course of the paper. In Fig. \ref{fig:photopol}(a), schematic configurations of X$^+$, X$^{+*}$, and X$^{+*}_p$ is shown together with a simplified picture of recombination processes associated with them (green arrows).

\begin{table}[t]
	\begin{ruledtabular}
		\caption{The amplitude of the FSS splitting measured for dot 1 as compare to its mean value deduced from experiments performed on 23 different QDs.}
		\label{tab:tab1}
		\begin{tabular}{lcc}
			& dot 1 & Ensemble \\
			\hline
			X$^{+*}$ & 130 \textmu eV & (118.8$\pm$2.3) \textmu eV \\
			X$^{+*}_p$ & 60 \textmu eV & (68.2$\pm$4.1) \textmu eV\\
		\end{tabular}
	\end{ruledtabular}
\end{table}

The PL spectra of the three characteristic lines detected for different directions of linear polarization are illustrated in Fig. \ref{fig:photopol}(b) and (c). Note that the X$^{+*}_p$ line appears in the emission spectra at lowest excitation power densities simultaneously with the X$^+$ and X$^{+*}$ lines. This is despite the fact that the X$^{+*}_p$ emission involves a transition within the $p$-shell. As shown in Fig.~\ref{fig:photopol}(b)~and~(c), the X$^+$ line is a single component line, while the X$^{+*}$ and X$^{+*}_p$ display two perpendicularly partially polarized components. The X$^+$ line is attributed to the recombination of the spin-singlet state of the charged exciton (the complex consists of an $s$-shell electron ($e_s$) and two $s$-shell holes ($2h_s$)).\cite{bayerprb}  The $eh$ exchange interaction influences neither the initial (in which two holes form a closed shell with the total spin $J_h=0$) nor the final state (only one hole left). As a result no fine structure splitting (FSS) appears for the X$^+$ line. The fine structures of the X$^{+*}$ and X$^{+*}_p$ lines are due to the splitting of the initial states. In both cases the final states consist of only one hole. For the X$^{+*}$, the spin-triplet configuration of the charged exciton, the initial state comprises an $s$-shell electron ($e_s$), an $s$-shell hole ($h_s$), and a $p$-shell hole ($h_p$). The two holes form a spin-singlet ($J_h=0$) and a spin-triplet ($J_h=3$) states split by the $hh$ exchange energy. In addition, the $eh$ exchange interaction between the respective holes and the electron $e_s$ splits the triplet states into three doublets.\cite{arashidatri} As a result the X$^{+*}$ line comprises two partially linearly polarized components oriented along two perpendicular crystallographic directions $[110]$ and $[1\bar{1}0]$.\cite{molascharged} The magnitude of the splitting equals about 130~\textmu eV. In the case of the X$^{+*}_p$ line, argued to be the spin-triplet configuration of the excited charged exciton, the origin of the FSS is analogous to the discussed for the X$^{+*}$ line. The initial state of the complex, which corresponds to this line is composed of a $p$-shell electron ($e_p$), an $s$-shell hole ($h_s$), and a $p$-shell hole ($h_p$). Similarly, the $hh$ and $eh$ exchange interactions lead to two partially linearly polarized components of  X$^{+*}_p$ with the splitting magnitude of approximately 60~\textmu eV.

The analysis of the FSS magnitude of the X$^{+*}$ and X$^{+*}_p$ lines was performed on many other single QDs present in the sample (23 in total). Results of this study are summarized in Tab. \ref{tab:tab1}. In average, the FSS of the X$^{+*}$ line is found to be of about two times bigger than for the X$^{+*}_p$ line. This difference is very likely associated to the variation in the strength of $eh$ exchange interaction in one case implying the electron from the $s$-shell and in another case from the $p$-shell, each shell having different spatial extent and symmetry.\cite{korkusinski}  Nevertheless, more theoretical studies are needed to clarify this point.

\begin{figure}[b!]
\centering
\includegraphics[width=60mm]{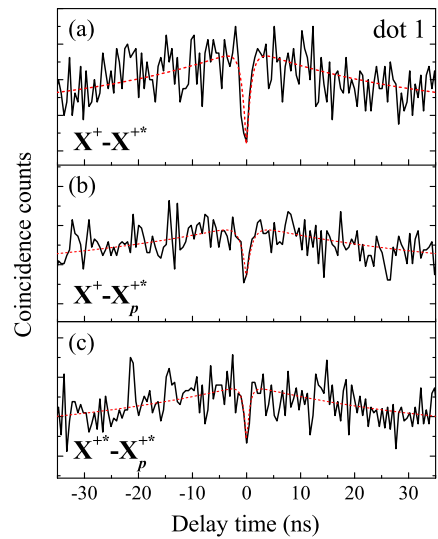}
\caption{The photon cross-correlation histogram of the X$^{+}$, X$^{+*}$, X$^{+*}_{p}$ emission lines. Dashed red lines are the guides to the eye.\label{fig:correlation}}
\end{figure}

Our assignment of the X$^+$, X$^{+*}$, and X$^{+*}_p$ states as due to different states (ground and excited) but of the same complex (positively charged exciton) is strongly supported by the results of photon correlation experiments. The histograms, which correspond to cross-correlation between the emission of investigated lines are presented in Fig. \ref{fig:correlation}. All histograms display the processes characteristic of rather long time scale (bunching-like peak), of the order of 10~-~20~ns, on the top of which short scale (<1~ns) processes (antibunching dip) are seen. The long-scale processes reflect the charge fluctuation effects (random capture of single carriers) in the QD under a quasi-resonant excitation (below the energy of the barrier).\cite{pietkaprb,nakajima} Sharp antibunching peaks around zero delay time (characteristic of processes on the scale of the radiative recombination time) indicate that at the same time photons are emitted either in the X$^+$, or in the X$^{+*}$, or in the X$^{+*}_p$ recombination process. This is in agreement with our assumption of three different recombination processes each leading to "disappearance" of the same (three-particle) object.

\begin{figure}[t]
\centering
\includegraphics[width=80mm]{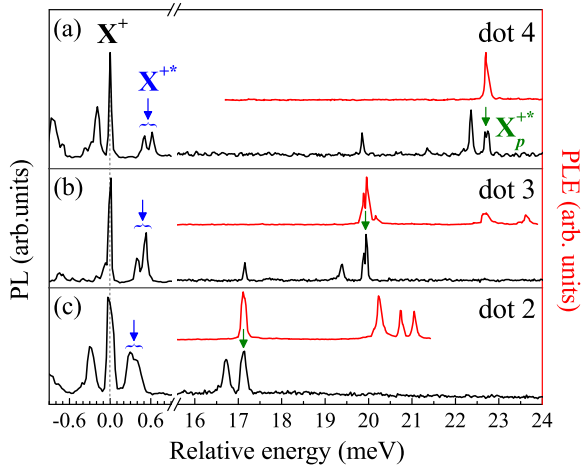}
\caption{(color online) The PL spectra (black lines) and the PLE spectra (red lines) detected on the X$^{+*}$ line of three different single QDs. The horizontal scale was shifted with reference to the energy of the X$^+$ line ($\mathrm{E}_{\mathrm{X}^{\mathrm{+}}}$= (a)  1.60483~eV, (b) 1.60347~eV, (c) 1.60214~eV). The spectra are normalized to the intensity of the X$^+$ line and vertically shifted for the sake of clarity.\label{fig:plplusplex3}}
\end{figure}

The relationship between X$^+$, X$^{+*}$, and X$^{+*}_p$ can be also seen when comparing the PL and PLE spectra of the investigated dots. This is demonstrated in Fig.~\ref{fig:plplusplex3} with pairs of PL and PLE spectra measured for three different dots. It can be observed that the PLE spectrum of the X$^{+*}$ emission shows a clear (absorption) resonance which perfectly coincides with the X$^{+*}_p$ emission peak, and the same is true for the PLE spectrum of the X$^+$ emission peak (data not shown). The observation of the X$^{+*}_p$ resonance in the PL spectra implies that the relaxation from the X$^{+*}_p$ is quite suppressed but on the other hand it remains efficient enough to give raise to the resonance in the PLE of X$^{+*}$ (and X$^{+}$) emission lines. Notably, the X$^{+*}_p$ resonance shows the characteristic FSS doublet structure both in PL and PLE (absorption) spectra. This is not surprising as the exchange effects are active in both the recombination and absorption processes, in their initial and final states, respectively.

\begin{figure}[t]
\centering
\includegraphics[width=60mm]{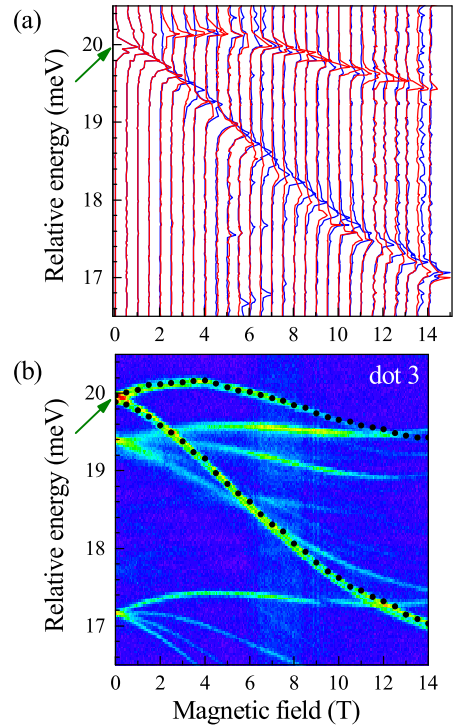}
\caption{(color online) (a) The PLE spectra detected on the X$^{+*}$ line as a function of the magnetic field. The red and blue traces indicate the $\sigma_{+}$- and $\sigma_{-}$-polarised components of the X$^{+*}$ line, on which the spectra are measured. The spectra are normalized to the to the most intense lines. (b) Magnetic-field evolution of the X$^{+*}_p$ resonance (black points), as extracted from the PLE spectra presented in part (a), superimposed on a color-scale map displaying the PL spectra related to the $p$-shell emission of the same dot, plotted as a function of the magnetic field. The green arrows indicate the energy position of the X$^{+*}_{p}$ line at 0 T. The vertical scale is shifted with reference to the energy of the X$^+$ line at B=0 T.\label{fig:plefield}}
\end{figure}

\begin{figure}[t]
	\centering
	\includegraphics[width=80mm]{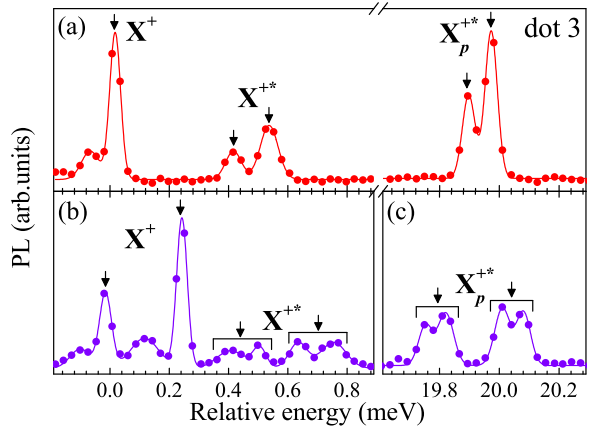}
	\caption{(color online) The PL spectra of a single GaAlAs QD measured at low excitation power at (a) 0~T, (b) 3~T, and (c) 1~T. The horizontal scale is shifted with reference to the energy of the X$^+$ line at B=0~T. ($\mathrm{E}_{\mathrm{X}^{\mathrm{+}}}$=1.60347~eV).
		\label{fig:PLmagn}}
\end{figure}

More can be learned about the properties of the X$^{+*}_{p}$ state from PL and PLE measurements performed in magnetic fields. A collection of data which demonstrate the evolution of the X$^{+*}_{p}$ resonance with magnetic field in both types of experiments is displayed in Fig.~\ref{fig:plefield} (for dot 3). 
Firstly, it is important to emphasize that the magnetic-field dependence of the X$^{+*}_p$ resonance obtained from PLE spectra follows exactly the same evolution as that of the X$^{+*}_p$ line in PL spectra (see Fig. \ref{fig:plefield}(b)). In the presence of magnetic field, the X$^{+*}_{p}$ resonance splits, at first sight into two well resolved components which center of mass displays the red shift with the magnetic field. The red shift of spectral lines is a fingerprint of resonances which involve $p$-shell electrons. This is in agreement with our assignment of the X$^{+*}_{p}$ state. The energy separation between magnetically split components is linear with magnetic field. The X$^{+*}$ resonance splits in magnetic field as well, but the amplitude of the X$^{+*}_{p}$ splitting is of about three times larger then the corresponding splitting of the X$^{+*}$ state. This can be seen in Fig. \ref{fig:PLmagn}, which depicts a similar magnetically-induced splitting of the X$^{+*}$ and X$^{+*}_{p}$ emission lines with the former measured at B=3~T and the latter at B=1~T. To this end, the data presented in Fig. \ref{fig:PLmagn} demonstrate that each magnetically split component consists in fact of a doublet, in both cases of the X$^{+*}$ and X$^{+*}_{p}$ resonances. 
The fine, doublet structure of the magnetically split components is a characteristic feature observed for the X$^{+*}$ resonance. The analogous effect observed for the X$^{+*}_{p}$ resonance confirms our hypothesis on the origin of this resonance.

We believe that altogether our experimental data provide convincing arguments for the identification of three characteristic spectral lines in the investigated QDs as due to three different X$^+$, X$^{+*}$, and X$^{+*}_p$ states of a positively charged exciton. Most intriguing are the properties of the X$^{+*}_{p}$ resonance which involves $p$-shell electrons and appears up to 20~meV above the X$^+$ and X$^{+*}$ states, both comprising the $s$-shell electrons. Its observation in the PL spectra implies the partial blocking of the relaxation of the X$^{+*}_{p}$ state towards lower energy X$^+$ and X$^{+*}$ states. A suppression of the relaxation processes between QD states is most often discussed in reference to spin flip transitions which are often speculated to be inefficient. The suppressed efficiency of the spin relaxation was indeed invoked to explain the appearance of the PL signal from the X$^{+*}$ resonance (the X$^{+*}$ to X$^+$ relaxation implies the change of the spin of the exciton and/or of a hole in a simplified view).\cite{benny2012,arashidatri,molascharged} Spin-blocking arguments could also account for the suppressed relaxation between X$^{+*}_p$ and X$^+$ but they do not hold in the case of the X$^{+*}_p$ to X$^{+*}$ transition which, in a simplified picture, does not imply the flip of spin. On the other hand the magnetic field splitting of the X$^{+*}_p$ resonance is much larger than that of the X$^{+*}$ (and X$^+$) states, which may suggest much complex picture of spin physics (coupling of spin and orbital momenta) associated with those states. Optionally, the observed reduced efficiency of relaxation from the $p$- to $s$-shell states might be due to the fact that the energy separation between these states is exceptionally small in the investigated dots. This separation is much smaller then any characteristic energy of optical phonon of the host lattice. Therefore, the discussed relaxation process must involve acoustic phonons, which are not necessarily well coupled to the electronic states.

Whatever is the reason of the suppressed relaxation in the investigated dots, we take this fact as an advantage and study yet another X$^{+*}_p$ resonance in a single QD. Although the overall properties of this state are quite similar to those exhibited by the X$^{+*}$ state, the amplitude of the effects such as fine structure and magnetic field splitting is very different for both these states. For the X$^{+*}_p$ state, the fine structure splitting is of about two-times smaller whereas the magnetic splitting is three times larger as compared to the case of the X$^{+*}$ state. We believe that this is due to different symmetry of $s$- and $p$-shell electrons involved in these two states, but more theoretical studies are needed to quantify those effects.

\section{Summary \label{sec:conclussions}}

Concluding, we have identified three different states of the positively charged exciton in a single GaAlAs/AlAs QD. Their properties have been studied using PL and PLE spectroscopy as well as photon correlation experiments. Linear polarization of the emission spectra has been analyzed to examine the fine structure splitting 
whereas the PL and PLE spectra measured as a function of magnetic field have been used to determine the Zeeman splitting of these states. The studied resonances are assigned to the singlet ground state (X$^+$) of positively charged exciton and its two excited triplet states (holes with the same spin): X$^{+*}$ involves the electron from the $s$-shell and X$^{+*}_p$ involves the electron from the $p$-shell state. The observation of highly excited X$^{+*}_p$ state in emission spectra implies a significant suppression of the relaxation processes in the dots. Particular attention has been paid to compare the properties of two triplet states which show quite different amplitude of the fine structure and Zeeman effects. Possible paths towards the interpretation of our observations are indicated, but those remain to be confronted with conceivable theoretical studies.

\section{Acknowledgements}
The work has been supported by the Foundation for Polish Science International PhD Projects Programme co-financed by the EU European Regional Development Fund. M.R.M. kindly acknowledges the National Science Center (Grants No. DEC-2013/08/T/ST3/00665 and DEC-2013/09/N/ST3\linebreak[4]/04237) for financial support for his PhD.

\bibliographystyle{apsrev4-1}
\bibliography{bibliomagPLE}

\end{document}